\documentclass[aps,prl,twocolumn,showpacs,floatfix,superscriptaddress]{revtex4}
\usepackage{dcolumn}
\usepackage{graphicx}
\input{psfig.sty}
\newcommand{\arxivonly}[1]{}  %% for the version to be sent to PRL

\begin{document}
% \linenumbers

%%\preprint{CDF/ANAL/JET/CDFR/7785}
%%\preprint{Version 3.0}

\title{Measurement of the Inclusive Jet Cross Section
in {\boldmath $p\bar p$} Interactions at {\boldmath $\sqrt{s}=1.96$} TeV
Using a Cone-based Jet Algorithm}

%%\author{A. Bhatti, F. Chlebana, G. Flanagan, K. Hatakeyama, J. Huston, G. Latino}
\affiliation{Institute of Physics, Academia Sinica, Taipei, Taiwan 11529, Republic of China} 
\affiliation{Argonne National Laboratory, Argonne, Illinois 60439} 
\affiliation{Institut de Fisica d'Altes Energies, Universitat Autonoma de Barcelona, E-08193, Bellaterra (Barcelona), Spain} 
\affiliation{Baylor University, Waco, Texas  76798} 
\affiliation{Istituto Nazionale di Fisica Nucleare, University of Bologna, I-40127 Bologna, Italy} 
\affiliation{Brandeis University, Waltham, Massachusetts 02254} 
\affiliation{University of California, Davis, Davis, California  95616} 
\affiliation{University of California, Los Angeles, Los Angeles, California  90024} 
\affiliation{University of California, San Diego, La Jolla, California  92093} 
\affiliation{University of California, Santa Barbara, Santa Barbara, California 93106} 
\affiliation{Instituto de Fisica de Cantabria, CSIC-University of Cantabria, 39005 Santander, Spain} 
\affiliation{Carnegie Mellon University, Pittsburgh, PA  15213} 
\affiliation{Enrico Fermi Institute, University of Chicago, Chicago, Illinois 60637} 
\affiliation{Joint Institute for Nuclear Research, RU-141980 Dubna, Russia} 
\affiliation{Duke University, Durham, North Carolina  27708} 
\affiliation{Fermi National Accelerator Laboratory, Batavia, Illinois 60510} 
\affiliation{University of Florida, Gainesville, Florida  32611} 
\affiliation{Laboratori Nazionali di Frascati, Istituto Nazionale di Fisica Nucleare, I-00044 Frascati, Italy} 
\affiliation{University of Geneva, CH-1211 Geneva 4, Switzerland} 
\affiliation{Glasgow University, Glasgow G12 8QQ, United Kingdom} 
\affiliation{Harvard University, Cambridge, Massachusetts 02138} 
\affiliation{Division of High Energy Physics, Department of Physics, University of Helsinki and Helsinki Institute of Physics, FIN-00014, Helsinki, Finland} 
\affiliation{University of Illinois, Urbana, Illinois 61801} 
\affiliation{The Johns Hopkins University, Baltimore, Maryland 21218} 
\affiliation{Institut f\"{u}r Experimentelle Kernphysik, Universit\"{a}t Karlsruhe, 76128 Karlsruhe, Germany} 
\affiliation{High Energy Accelerator Research Organization (KEK), Tsukuba, Ibaraki 305, Japan} 
\affiliation{Center for High Energy Physics: Kyungpook National University, Taegu 702-701; Seoul National University, Seoul 151-742; and SungKyunKwan University, Suwon 440-746; Korea} 
\affiliation{Ernest Orlando Lawrence Berkeley National Laboratory, Berkeley, California 94720} 
\affiliation{University of Liverpool, Liverpool L69 7ZE, United Kingdom} 
\affiliation{University College London, London WC1E 6BT, United Kingdom} 
\affiliation{Massachusetts Institute of Technology, Cambridge, Massachusetts  02139} 
\affiliation{Institute of Particle Physics: McGill University, Montr\'{e}al, Canada H3A~2T8; and University of Toronto, Toronto, Canada M5S~1A7} 
\affiliation{University of Michigan, Ann Arbor, Michigan 48109} 
\affiliation{Michigan State University, East Lansing, Michigan  48824} 
\affiliation{Institution for Theoretical and Experimental Physics, ITEP, Moscow 117259, Russia} 
\affiliation{University of New Mexico, Albuquerque, New Mexico 87131} 
\affiliation{Northwestern University, Evanston, Illinois  60208} 
\affiliation{The Ohio State University, Columbus, Ohio  43210} 
\affiliation{Okayama University, Okayama 700-8530, Japan} 
\affiliation{Osaka City University, Osaka 588, Japan} 
\affiliation{University of Oxford, Oxford OX1 3RH, United Kingdom} 
\affiliation{University of Padova, Istituto Nazionale di Fisica Nucleare, Sezione di Padova-Trento, I-35131 Padova, Italy} 
\affiliation{University of Pennsylvania, Philadelphia, Pennsylvania 19104} 
\affiliation{Istituto Nazionale di Fisica Nucleare Pisa, Universities of Pisa, Siena and Scuola Normale Superiore, I-56127 Pisa, Italy} 
\affiliation{University of Pittsburgh, Pittsburgh, Pennsylvania 15260} 
\affiliation{Purdue University, West Lafayette, Indiana 47907} 
\affiliation{University of Rochester, Rochester, New York 14627} 
\affiliation{The Rockefeller University, New York, New York 10021} 
\affiliation{Istituto Nazionale di Fisica Nucleare, Sezione di Roma 1, University of Rome ``La Sapienza," I-00185 Roma, Italy} 
\affiliation{Rutgers University, Piscataway, New Jersey 08855} 
\affiliation{Texas A\&M University, College Station, Texas 77843} 
\affiliation{Istituto Nazionale di Fisica Nucleare, University of Trieste/\ Udine, Italy} 
\affiliation{University of Tsukuba, Tsukuba, Ibaraki 305, Japan} 
\affiliation{Tufts University, Medford, Massachusetts 02155} 
\affiliation{Waseda University, Tokyo 169, Japan} 
\affiliation{Wayne State University, Detroit, Michigan  48201} 
\affiliation{University of Wisconsin, Madison, Wisconsin 53706} 
\affiliation{Yale University, New Haven, Connecticut 06520} 
\author{A.~Abulencia}
\affiliation{University of Illinois, Urbana, Illinois 61801}
\author{D.~Acosta}
\affiliation{University of Florida, Gainesville, Florida  32611}
\author{J.~Adelman}
\affiliation{Enrico Fermi Institute, University of Chicago, Chicago, Illinois 60637}
\author{T.~Affolder}
\affiliation{University of California, Santa Barbara, Santa Barbara, California 93106}
\author{T.~Akimoto}
\affiliation{University of Tsukuba, Tsukuba, Ibaraki 305, Japan}
\author{M.G.~Albrow}
\affiliation{Fermi National Accelerator Laboratory, Batavia, Illinois 60510}
\author{D.~Ambrose}
\affiliation{Fermi National Accelerator Laboratory, Batavia, Illinois 60510}
\author{S.~Amerio}
\affiliation{University of Padova, Istituto Nazionale di Fisica Nucleare, Sezione di Padova-Trento, I-35131 Padova, Italy}
\author{D.~Amidei}
\affiliation{University of Michigan, Ann Arbor, Michigan 48109}
\author{A.~Anastassov}
\affiliation{Rutgers University, Piscataway, New Jersey 08855}
% ...
\collaboration{CDF Collaboration}
\noaffiliation

\date{\today}

\begin{abstract}
We present a measurement of the inclusive jet cross section in
$p{\bar p}$ interactions at $\sqrt{s}=1.96$ TeV %%.
using 385~${\rm pb}^{-1}$ of data 
collected with the CDF II detector at the Fermilab Tevatron.
The results are obtained
using an improved cone-based jet algorithm (Midpoint). The data cover the jet 
transverse momentum range from 61 to 620 GeV/{\it c}, extending the reach by 
almost 150 GeV/{\it c} compared with previous measurements at the Tevatron.
The results 
are in good agreement with next-to-leading order perturbative QCD predictions using the
CTEQ6.1M parton distribution functions. 
\end{abstract}

\pacs{13.87.Ce, 12.38.Qk, 13.85.Ni}

\maketitle
%%%%%%%%%%%%%%%%%%%%%%%%
\clearpage

The differential jet production cross section at the Tevatron probes the world's highest 
momentum transfers in particle collisions, is potentially sensitive to a wide variety of 
new physics, such as quark compositeness~\cite{b:compositness}, and tests perturbative QCD (pQCD) over more 
than eight orders  of magnitude.
There was great interest when the inclusive jet cross section measured by the CDF collaboration 
at the center of mass energy 
$\sqrt{s}=1.8$ TeV ~\cite{b:CDFRunIPRL,b:CDFRunIPRD} exhibited an excess in the 
high transverse energy
$E_T$ region when compared to next-to-leading order (NLO) QCD predictions obtained using 
then-current parton distribution 
functions (PDFs)~\cite{b:cteq4m}.  
Global PDF analysis by CTEQ~\cite{b:cteq4hj,b:cteq5hj} demonstrated that  
the excess could be explained by an enhanced gluon distribution at
high momentum fraction $x$ ($x>0.3$).
Recent global PDF fits (CTEQ6, CTEQ6.1, MRST2004)~\cite{b:cteq6m,b:cteq6_jet,b:mrst2004},
which include the Run I Tevatron jet data~\cite{b:CDFRunIPRL,b:d0RunI}, 
find an increased gluon density at high $x$ and provide a good 
description of the Run I Tevatron data.
The gluon distribution is still poorly constrained 
%%in this $x$ region
at high $x$ (see e.g. Ref.\cite{b:cteq6_jet}) and contributes 
significantly to the theoretical uncertainty for many interesting processes 
at the Tevatron and the LHC. 
The increase in $\sqrt{s}$ from 1.8 to 1.96~TeV, together with higher luminosity in 
Run~II, allows more precise jet production measurements with a significantly extended kinematic 
range.

Jet algorithms cluster together
objects such as partons or particles, or energies measured in
calorimeter cells.  The clustering algorithms rely on the
association of these objects based on transverse momenta  (the $k_T$  
algorithm), or
angles (the cone algorithm), relative to a jet axis.
A measurement using the $k_T$ algorithm is reported in
Ref.~\cite{b:kT_prl}.
In this letter, we report the results of an inclusive jet measurement using a 
cone algorithm for the rapidity region 
$0.1<|y|<0.7$~\cite{b:coordinate}.
%%The analysis is described in greater detail in Ref.~\cite{b:gene}.
Cone jet algorithms, rather than $k_T$ algorithms,  have been used dominantly 
at hadron collider experiments mainly due to the simplicity in 
constructing corrections for the underlying event and for multiple interactions in
the same bunch crossing~\cite{b:RunII_workshop}.
It is worth noting that, previously, results from a cone
algorithm~\cite{b:d0RunI} and $k_T$ algorithm~\cite{b:d0RunIKt} by the
D\O\ collaboration showed only marginal agreement at low $p_T$ where
corrections for multiple interactions and underlying event become
important. %%, and
%%it implies the importance of measuring the inclusive jet cross section
%%with a cone algorithm as well.
We use the Midpoint algorithm, an improved iterative cone clustering 
algorithm~\cite{b:RunII_workshop}. 
It is difficult to use previous iterative cone
algorithms~\cite{b:CDFRunIPRD,b:d0RunI} with higher 
order pQCD calculations due to the presence of infrared singularities.
The Midpoint algorithm places additional seeds 
%%KH (directions for jet cones)
between stable cones having a
separation of less than twice the size of the clustering cones; the use 
of these additional seeds reduces the problem with infrared
singularities.
%%KH The Midpoint algorithm uses 
%%KH four-vector kinematics for clustering individual partons, particles, or 
%%KH energies measured in calorimeter cells, and 
%%KH jets are described using 
%%KH rapidity ($y$) and transverse momentum ($p_T$);
%%KH the CDF Run I jet clustering algorithm used scalar summation for
%%KH clustering, pseudorapidity ($\eta$)~\cite{b:coordinate},
%%KH and the transverse energy ($E_T$) as described in Ref.~\cite{b:Snowmass}.

%describe the CDF detector 
%\section{detector description}

The CDF II detector is a magnetic spectrometer which is described in detail 
elsewhere~\cite{b:CDFIITDR}. Here we describe briefly those components
that are crucial to this measurement. 
The central detector consists of a silicon vertex detector 
inside a cylindrical drift chamber. Surrounding the tracking detectors is a 
superconducting solenoid  which provides a 1.4 T magnetic field. 
Outside the solenoid is the central calorimeter, covering a
pseudorapidity ($\eta$)~\cite{b:coordinate} range up to 1.1.
The central calorimeter consists of 48 modules, segmented into towers of granularity 
$\Delta \eta \times \Delta\phi \approx  0.1 \times 0.26$ and divided 
into electromagnetic (CEM) and hadronic (CHA) sections. 
The CEM is a lead-scintillator calorimeter;
the CHA  is an iron-scintillator calorimeter with a depth of approximately 
4.7 interaction lengths. 
The energy resolution of the CEM for electrons is 
$\sigma(E_T) / E_T = 13.5\%/\sqrt{E_T ({\rm GeV})} \oplus 2\%$ while the average
energy resolution of the CHA for charged pions is 
$\sigma(E_T) / E_T = 50\%/\sqrt{E_T ({\rm GeV})} \oplus 3\%$. 
The forward region, $1.1 < |\eta| < 3.6$,  is covered by the ``Plug Calorimeters''
consisting of lead-scintillator for the electromagnetic section 
and iron-scintillator for the hadronic section. The region between the 
central and forward calorimeters, $0.7 < |\eta| <1.3$, is covered by an iron-scintillator hadron
calorimeter with similar segmentation to the central calorimeter.

%\section{Event Selection}

This measurement uses a data sample corresponding to an integrated
luminosity of 385~${\rm pb}^{-1}$ collected between February
2002 and August 2004.
The data were collected using four trigger paths.
The Level 1 trigger requires a calorimeter trigger tower, consisting of
two calorimeter towers adjacent in $\eta$, to have either $E_T>5$ GeV 
or $E_T>10$ GeV. At Level 2, the calorimeter towers are clustered using 
a nearest neighbor algorithm. Four trigger paths with cluster $E_T >$ 15, 40, 60, and 90 GeV
are used. 
Events in these paths are required to pass jet $E_T >$ 20 (J20), 50 (J50), 70 (J70), 
and 100 (J100) GeV thresholds at Level 3, where the clustering is
%%KH Events in these paths are required to pass jet $E_T >$ 20, 50, 70, 
%%KH and 100 GeV thresholds at Level 3, where the clustering is
performed using a cone algorithm with a cone radius $R_{cone}=0.7$.

Cosmic ray events are rejected by a cut on the missing transverse energy 
(${\not\!\!E_T}$) significance~\cite{b:missingEt}.
For J20, J50, J70, and J100, we remove events having a ${\not\!\!E_T}$ significance 
greater than 4, 5, 5, and 6 GeV$^{1/2}$, respectively. 
In addition, all events containing 
jets with $p_T>360$ GeV/{\it c} and passing the analysis cuts were
visually scanned, and no cosmic ray events were found.
The efficiency of the ${\not\!\!E_T}$ significance cut is 100\% for jets
at low $p_T$ (65 GeV/{\it c}) and decreases to 92~\% for jets at high
$p_T$ (550 GeV/{\it c}).
We reconstruct $z$-vertices by fits to tracks in the event and
a beamline constraint, and select the vertex with the highest total
$p_T$ of the associated tracks as the event vertex.
%%This finds the correct event vertex greater than 99.5\% of the time.
In order to ensure that the jet energy is well measured, the event 
vertex is required to be within 60 cm of the center of the 
detector along the beamline. %% direction.
The efficiency for this cut is determined to be 95\% from the beam
profile measured using a minimum bias sample. 
Jets are required to have a rapidity $|y|$ between 0.1 and 0.7 
to reduce the effects of the gap between calorimeter modules and at the 
transition region between the central and plug calorimeters. 

There are two essential stages for any jet algorithm. First, the objects 
(partons, particles, or calorimeter towers) 
belonging to a  cluster are identified.  With the Midpoint algorithm
the cluster is a cone of radius 0.7 in $(y,\phi)$ space. For reasons
dealing with problems of unclustered energy
endemic to iterative cone algorithms~\cite{b:building},
the clustering radius is at first set to
$R_{cone}/2 (=0.35)$, and then later expanded to its full size as discussed
below.
%The clustering algorithm itself is divided into two stages. In the 
%first stage, the cluster  radius is set  to $R_{cone}/2$. This step is 
%necessary to reduce problems with unclustered  energy endemic to cone 
%jet  algorithms ~\cite{b:building}.
Second, the kinematic variables
defining the jet are calculated from the objects comprising a cluster.
The Midpoint algorithm makes use of four-vectors throughout the clustering process. 
The four-vector for each tower is computed as a sum of vectors
for the electromagnetic and hadronic compartments of the tower;
the vector for each compartment is defined by assigning a mass-less
vector with magnitude equal to the deposited energy and with direction
from the event vertex to the center of each compartment~\cite{b:RunII_workshop}.
The detector towers are sorted in order of descending $p_T$. Only towers  passing a seed cut, 
$p_T^{tower}>p_T^{seed}$, are used as starting points for the initial jet cones.
The seed threshold is chosen to be 1 GeV/{\it c}; its value has 
a negligible effect on jets in the kinematic region used in this measurement.
A tower passing the threshold of 100 MeV/{\it c} is clustered into a
cone and eventually into a jet if the separation from the axis of the
cone in $(y,\phi)$ is smaller than $R_{cone}/2$.
There is no threshold for particle and parton clustering. 
%For this measurement, we use a cone radius of 0.7 defined in
%$(y,\phi)$ space.
After each iteration the jet centroid position is updated. 
The jet clustering is 
repeated until all of the jet cones are stable. A cone is stable when the
tower list is unchanged from the previous iteration. %%from one iteration to the next. 
%After all stable cones have been 
%determined, an additional seed is 
%defined at the midpoint between any two cones 
%separated by less than twice the clustering radius, and the process is repeated.
%At this point, the cluster  radius is expanded to the full size 
%($R_{cone}$).
%%JH added the sentence below
After all stable cones have been determined, the clustering radius is expanded to the full size ($R_{cone}$). The use of the smaller initial cone results in an expected cross section approximately 5\% larger due to the inclusion of jet energy that would have remained unclustered in the default Midpoint algorithm~\cite{b:RunII_workshop}.
At this point, an additional seed is defined at the midpoint between any two cones separated
by less than $2R_{cone}$ and the iteration process is repeated.
Two overlapping cones, if present, are merged into a single jet if the shared energy
is larger than $75\%$ ($f_{merge} = 0.75$) of the energy of the jet with 
lower $p_T$; otherwise the shared towers are
 assigned to the nearest
jet. This splitting/merging procedure is iterated until the tower
assignments to jets are stable.
The jet kinematic properties are defined using a four-vector recombination scheme~\cite{b:RunII_workshop}.
%KH as (four-vector recombination scheme):
%
%KH \begin{eqnarray}
%KH p^{jet}&=&(E^{jet},{\bf p}^{jet})=\sum_{i}(E^i,p_x^i, p_y^i, p_z^i ), \nonumber \\
%KH p_{_{T}}^{jet}&=&\sqrt{{(p_x^{jet}})^2 + ({p_y^{jet}})^2},   \nonumber  \\
%KH y^{jet}&=& \frac{1}{2}\ln \left (\frac{E^{jet}+p_z^{jet}}{E^{jet}-p_z^{jet}}\right ),       \nonumber  \\
%KH \phi^{jet} &=& \tan^{-1}\left (\frac{p_y^{jet}}{p_x^{jet}}\right ).  
%KH \end{eqnarray}
%

The inclusive differential jet cross section is defined as:
\begin{equation}
\frac{ d^2\sigma}{dp_T dy } 
= \frac{ 1}{\Delta y}\frac{ 1}{\int L dt}\frac{N_{jet}/\epsilon }{\Delta p_T },
\end{equation} 
where $N_{jet}$ is the number of jets in the $p_T$ range $\Delta p_T$, 
$\epsilon$ is the trigger, ${\not\!\!E_T}$ significance cut and
$z$-vertex cut efficiency, $\int L dt$ 
is the effective integrated luminosity, and $\Delta y = 1.2$ is the 
rapidity interval used in the 
analysis. A trigger efficiency greater than $99.5\%$ is required to include the 
jets collected by a given trigger threshold.
The measured calorimeter level jet cross section must be corrected for detector effects and 
for energy from additional
$p\bar p$ interactions in the same bunch crossing (pile-up). 
For this sample, the average number of additional $p\bar p$
interactions is about 0.9. 
The pile-up corrections subtract $0.93(\pm0.14)$ GeV/{\it c} for each
additional $z$-vertex from the measured jet
$p_T$~\cite{b:JetCorNIM}.

The detector response corrections are determined from a detector
simulation and a jet fragmentation model. 
The detector response is determined using a {\sc geant}-based detector 
simulation~\cite{b:cdfsim} 
in which a parametrized shower simulation
({\sc gflash}~\cite{b:gflash}) is used for the calorimeter simulation.
The {\sc gflash} parameters are tuned to test-beam data for electrons
and high-$p_T$ charged pions and to the collision data for low-$p_T$
charged hadrons~\cite{b:JetCorNIM}.
{\sc pythia} 6.216~\cite{b:pythia}, with Tune A~\cite{b:TuneA,b:jetshape},
is used for the production and fragmentation of the jets. 
Tune A refers to the values of the
parameters describing multiple-parton interactions and initial state radiation 
which have been adjusted to reproduce the energy observed in the
region transverse to the leading jet in the data from Run I.
It has also been shown to provide a reasonable description of the
measured energy distribution inside a jet~\cite{b:jetshape}.

The measured $p_T$ spectrum must be corrected for detector effects before
it can be compared to theoretical predictions. 
We cluster the final state stable particles~\cite{b:cluster} in {\sc pythia} using the 
same algorithm as the one used to cluster calorimeter towers.  
The resulting jets contain all the particles from the $p\overline{p}$ collision, 
including those from the hard scatter,  multiple parton-parton interactions
and beam remnants. The correction, done in two correlated steps,
is determined from a large sample of {\sc pythia} events, passed through 
the CDF detector simulation. First, a $p_T$-dependent correction is determined 
by matching the particle jet to the corresponding calorimeter jet and is applied 
to each measured jet. 
A binned spectrum is formed from the corrected $p_T$ of each jet. 
The bin widths are chosen commensurate with jet energy resolution and statistics.
The $p_T$ correction ranges from 1.17 at low $p_T$ (65~${\rm GeV}/{\it c}$) to 1.04 at 
high $p_T$ (550~${\rm GeV}/{\it c}$). 
The spectrum must be further corrected for bin-to-bin jet migration due to 
the finite energy resolution of the calorimeter. This unfolding correction depends 
on the detector energy resolution and the true spectrum as well as the $p_T$-dependent correction 
that was applied in the first step. 
The {\sc pythia} events are reweighted to match the experimental spectrum 
before the correction factors are calculated. A bin-by-bin unfolding correction is then 
determined by taking the ratio of the binned hadron level cross section and calorimeter
level cross section corrected by the $p_T$-dependent correction.
%%The size of the unfolding correction varies from 1.26 at low $p_T$ to 
%%2.36 at high $p_T$. The applied corrections remove the detector 
The size of the unfolding correction varies from 1.30 at low $p_T$ to 
2.31 at high $p_T$. 
The applied corrections remove the detector 
effects from the raw cross section and the corrected hadron level cross section 
can now be compared to theoretical predictions. 

The main systematic uncertainties on the measured inclusive jet cross section arise 
from four sources: the jet energy scale, the jet energy resolution, the unfolding 
of the measured cross section to the hadron level cross section, and the luminosity.  
The dominant source of uncertainty is from the jet energy scale. The energy scale is 
known to better than 3\% over the entire 
transverse momentum range, leading to an uncertainty on  the jet cross section varying 
from 10\% at low $p_T$ 
%%KH to 60\% at high $p_T$. 
to ${}^{+58}_{-39}$\% at high $p_T$,
comparable to the uncertainty achieved by CDF in Run I.
%%This is better than the uncertainty achieved by CDF in Run I.
The uncertainty due to the jet $p_T$ resolution is determined by the $p_T$ resolution difference
between the data and the {\sc pythia} Monte Carlo. The uncertainty on the cross section 
varies from about 6\% at low $p_T$ to about 10\% at high $p_T$.
The uncertainty associated with the unfolding correction is determined by correcting a
{\sc herwig} 6.5~\cite{b:Herwig}
dijet sample using the corrections derived from the 
{\sc pythia} sample. This uncertainty is determined
to be less than 5\% at high $p_T$ and lesss than 10\% at low $p_T$. 
The luminosity uncertainty is 6\%, independent of $p_T$ and is not included in the 
quoted systematic error. 
Other effects considered were determined to have a negligible effect on the cross section.
Adding all of these contributions in quadrature yields a total
experimental systematic uncertainty on the inclusive jet cross section
varying from approximately 15\% at low $p_T$ to approximately %%60\% at high $p_T$.
${}^{+60}_{-40}$~\% at high $p_T$.
%%KH The total experimental systematic uncertainty on 
%%KH the inclusive jet cross section varies from approximately 15\% at  low $p_T$ 
%%KH to approximately %%60\% at high $p_T$.
%%KH ${}^{+60}_{-40}$~\% at high $p_T$.

\begin{figure}[!th]
\includegraphics[width=1.00\linewidth,clip=true]{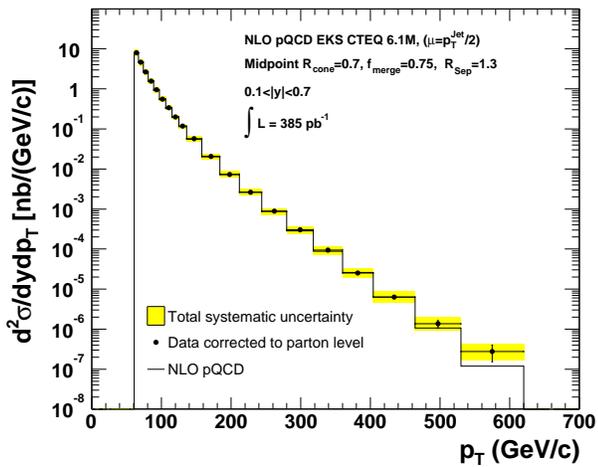}
\caption{\label{f:R2Log}
The measured inclusive jet differential cross section corrected to the parton level compared to the NLO pQCD
prediction of the EKS calculation using CTEQ6.1M.
}
\end{figure}

To compare the data with predictions for jets of partons as obtained from NLO 
calculations,
the data must be further corrected for underlying event and hadronization effects. 
%%JH added sentence below
It is also possible to correct the NLO predictions for the same
effects; the two corrections are simply the inverse of each other. 
For the former, 
we correct for the energy in the jet cone not associated with the hard scatter, i.e., 
from the collisions of other partons in the proton and antiproton. The latter corrects for 
particles outside the jet cone originating from partons whose 
trajectories lie inside the jet cone. It does not correct for the effects of 
hard gluon emission outside the jet cone, which are already accounted for in the 
NLO prediction.
The bin-by-bin hadron-to-parton corrections are obtained by applying the Midpoint clustering 
algorithm to the hadron level and to the parton level outputs of the
{\sc pythia} Tune A dijet Monte Carlo samples, generated with and without an underlying event. 
The sample without the underlying event was generated by turning off multiple parton 
interactions.
The underlying event correction results in a decrease of the cross section 
varying from 22\% at low $p_T$ to 4\% at high $p_T$; the hadronization correction 
increases the cross section by 13\% at low $p_T$, and by 3.5\% at high $p_T$. 
{\sc herwig} provides consistent results on
the hadronization corrections, but predicts smaller underlying event
energy; the difference in the underlying event correction 
%%KH which is
%%KH about 13\% at low $p_T$ and decreases with increasing $p_T$ 
is taken as the underlying event correction uncertainty.
In previous measurements at the Tevatron~\cite{b:CDFRunIPRL, b:d0RunI}, 
the hadronization corrections were not applied to the data.
%%\arxivonly{
The inclusive jet cross section is shown in Fig.~\ref{f:R2Log}, and 
Table~\ref{t:XSec1} lists the cross sections with the statistical and
systematic uncertainties at the hadron and parton levels. 
Also included in Table~\ref{t:XSec1} are the explicit factors applied to the hadron level 
cross section to obtain the parton level cross section. 
The experimental and theoretical jet cross sections are obtained by 
averaging over the transverse momentum bins. 
%%}

%\section{Comparison with QCD Predictions}

\begin{figure}[th]
\includegraphics[width=1.00\linewidth,clip=true]{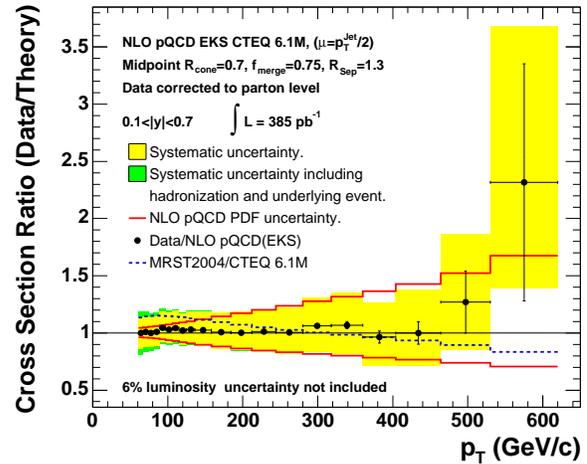}
\caption{\label{f:R2Lin1}
The ratio of the data corrected to the parton level over the NLO pQCD 
prediction of the EKS calculation using CTEQ6.1M. Also shown are the 
experimental systematic errors 
and the theoretical errors from the PDF uncertainty. 
The ratio of MRST2004/CTEQ6.1M is shown as the dashed line. An additional 
6\% uncertainty on the determination of the luminosity is 
not shown.
}
\end{figure}

Current NLO theoretical predictions for inclusive jet production exist only at the 
parton level, for which the final state consists only of 2 or 3 partons~\cite{b:eks,b:jetrad,b:jet++}.
For our comparisons with theory we use the calculation of EKS~\cite{b:eks}.
The ratio of the inclusive jet cross section, corrected to the parton level, to the 
NLO QCD predictions using the CTEQ6.1M PDFs is shown in Fig.~\ref{f:R2Lin1}. 
The Midpoint jet algorithm has been applied to the 2 or 3 partons in the final state of the EKS 
calculation. 
In order to mimic the properties of the splitting/merging step, present at the 
experimental level but not at the NLO 
parton level, a parameter $R_{sep}$ with a value of 1.3, has been introduced~\cite{b:building}. 
Two partons are clustered within the same jet if (1) they are within $R_{cone}$ (0.7 for this measurement) of the jet centroid and 
(2) within $R_{sep} \times R_{cone}$ of each other.  
The use of  $R_{sep}=1.3$  results in a reduction of the theoretical cross section 
prediction by approximately 5\%, roughly independent of jet transverse
momentum, as compared to the prediction obtained when $R_{sep}$ is not
used in the calculation.
In the EKS predictions, the renormalization and
factorization scales ($\mu_R$ and $\mu_F$) have both been set to $p_T^{jet}/2$.
Using a scale of $p_T^{jet}$ ($2p_T^{jet}$) rather 
than $p_T^{jet}/2$ leads to a theoretical prediction for the jet cross section 
lower by approximately 10\% (20\%) over the entire $p_T$ range and a larger 
$\chi^2$ in the global PDF fits~\cite{b:cteq6m}. The gluon distribution has been 
determined in the global fits, 
primarily by the Tevatron Run I jet data, using a renormalization and
factorization scale of $p_T^{jet}/2$. Thus, for self-consistency, this
scale should be used in the NLO comparisons. 

We show in Fig.~\ref{f:R2Lin1} the experimental uncertainties for the inclusive 
jet cross section 
and the theoretical uncertainties estimated from the 40 CTEQ6.1M error PDFs~\cite{b:cteq6m}. 
The PDF uncertainty is the dominant theoretical uncertainty for most of the transverse 
momentum range. The correction for underlying event and hadronization
is model dependent. The error associated with this correction is added in quadrature to the 
total experimental error and shown in Fig.~\ref{f:R2Lin1} as the outer shaded band. 
The data are in good agreement with the NLO QCD predictions,
which is consistent with what is reported in Ref.~\cite{b:kT_prl}.

It is important to emphasize that the CTEQ6.1M gluon density is already ``enhanced'' 
at high $x$ and so
automatically leads to a larger prediction for the jet cross section
than older PDFs such as CTEQ5M. 
Also shown in Fig.~\ref{f:R2Lin1} is the prediction using 
the latest PDF set from the MRST group~\cite{b:mrst2004}. The MRST2004 PDFs also contain 
an enhanced higher $x$ gluon, leading to reasonable agreement with the CDF jet measurement.

In conclusion, we have measured the inclusive jet cross section in the range
 $61 < p_T < 620$ GeV/{\it c}  
using an improved iterative cone clustering algorithm, Midpoint. The new measurement
extends the jet transverse momentum range over previous measurements at the Tevatron
by about 150 GeV/{\it c}. The data are well described by 
NLO QCD predictions using  CTEQ6.1M PDFs, within the theoretical (PDF) and 
experimental uncertainties.
%%JH added sentence below
 No new physics is indicated in the high $p_T$ region. 
Inclusion of these data in future global PDF fits will provide 
further constraints on the gluon distribution at large $x$.

We thank the Fermilab staff and the technical staffs of the
participating institutions for their vital contributions. This work
was supported by the U.S. Department of Energy and National Science
Foundation; the Italian Istituto Nazionale di Fisica Nucleare; the
Ministry of Education, Culture, Sports, Science and Technology of
Japan; the Natural Sciences and Engineering Research Council of
Canada; the National Science Council of the Republic of China; the
Swiss National Science Foundation; the A.P. Sloan Foundation; the
Bundesministerium f\"ur Bildung und Forschung, Germany; the Korean
Science and Engineering Foundation and the Korean Research Foundation;
the Particle Physics and Astronomy Research Council and the Royal
Society, UK; the Russian Foundation for Basic Research; the Comisi\'on
Interministerial de Ciencia y Tecnolog\'{\i}a, Spain; in part by the
European Community's Human Potential Programme under contract
HPRN-CT-2002-00292; and the Academy of Finland.

%%\arxivonly{
\begin{table*}[th]
\caption[Inclusive jet cross section corrected to parton level]{\label{t:XSec1}
Results for the inclusive jet cross section corrected to the 
hadron level, $d^2 \sigma^{hadron} / d p_T d y$,
and to the parton level, $d^2 \sigma^{parton} / d p_T d y$ are shown for each bin
together with the statistical (first) and systematic (second) errors. 
The correction factors, $C^{h\to p}$, applied to the hadron level
cross section to obtain the parton level cross section are also shown.
%%$C^{UE}$ is the underlying event correction and $C^{Had}$ is the
%%hadronization correction which
%%is applied to the data. The total hadron to parton level correction is given by
%%$C^{h \rightarrow p} = C^{UE} \cdot C^{Had}$.
%%The systematic error of $\frac{d^2 \sigma^{parton}}{d p_T d y}$
%%is the quadratic sum of the total experimental systematic error and 
%%the underlying event correction error.
There is an additional 6\% luminosity uncertainty.}
\begin{ruledtabular}
\begin{tabular}{cccc}
$p_T$  & $\frac{d^2 \sigma^{hadron}}{d p_T d y}$ & ~~$C^{h\to p}$~~ & 
$\frac{d^2 \sigma^{parton}}{d p_T d y}$  \\
~~(GeV/{\it c})~~    & (nb/(GeV/{\it c}))                     &          & (nb/(GeV/{\it c}))   \\ \hline
 
     61-67 &   $( 9.03 \pm  0.09 \stackrel{+ 1.26}{_{- 1.20} } )\times 10^{ 0}$ &   $ 0.889  \pm  0.008 \pm  0.116 $ &  $( 8.02 \pm  0.11  \stackrel{+ 1.53}{_{- 1.49} } )\times 10^{ 0}$ \\
     67-74 &   $( 5.17 \pm  0.05 \stackrel{+ 0.70}{_{- 0.65} } )\times 10^{ 0}$ &   $ 0.903  \pm  0.008 \pm  0.104 $ &  $( 4.67 \pm  0.06  \stackrel{+ 0.83}{_{- 0.80} } )\times 10^{ 0}$ \\
     74-81 &   $( 2.92 \pm  0.03 \stackrel{+ 0.39}{_{- 0.35} } )\times 10^{ 0}$ &   $ 0.916  \pm  0.009 \pm  0.092 $ &  $( 2.67 \pm  0.04  \stackrel{+ 0.45}{_{- 0.42} } )\times 10^{ 0}$ \\
     81-89 &   $( 1.70 \pm  0.02 \stackrel{+ 0.23}{_{- 0.20} } )\times 10^{ 0}$ &   $ 0.927  \pm  0.009 \pm  0.082 $ &  $( 1.57 \pm  0.02  \stackrel{+ 0.26}{_{- 0.23} } )\times 10^{ 0}$ \\
     89-97 &   $( 1.02 \pm  0.01 \stackrel{+ 0.14}{_{- 0.12} } )\times 10^{ 0}$ &   $ 0.936  \pm  0.007 \pm  0.073 $ &  $( 0.95 \pm  0.01  \stackrel{+ 0.15}{_{- 0.13} } )\times 10^{ 0}$ \\
    97-106 &   $( 5.90 \pm  0.04 \stackrel{+ 0.83}{_{- 0.69} } )\times 10^{-1}$ &   $ 0.945  \pm  0.007 \pm  0.064 $ &  $( 5.57 \pm  0.05  \stackrel{+ 0.87}{_{- 0.75} } )\times 10^{-1}$ \\
   106-115 &   $( 3.53 \pm  0.02 \stackrel{+ 0.51}{_{- 0.42} } )\times 10^{-1}$ &   $ 0.952  \pm  0.007 \pm  0.057 $ &  $( 3.36 \pm  0.03  \stackrel{+ 0.53}{_{- 0.44} } )\times 10^{-1}$ \\
   115-125 &   $( 2.07 \pm  0.01 \stackrel{+ 0.31}{_{- 0.25} } )\times 10^{-1}$ &   $ 0.958  \pm  0.007 \pm  0.050 $ &  $( 1.98 \pm  0.02  \stackrel{+ 0.31}{_{- 0.26} } )\times 10^{-1}$ \\
   125-136 &   $( 1.23 \pm  0.01 \stackrel{+ 0.19}{_{- 0.15} } )\times 10^{-1}$ &   $ 0.963  \pm  0.007 \pm  0.044 $ &  $( 1.18 \pm  0.01  \stackrel{+ 0.19}{_{- 0.16} } )\times 10^{-1}$ \\
   136-158 &   $( 5.84 \pm  0.03 \stackrel{+ 0.94}{_{- 0.76} } )\times 10^{-2}$ &   $ 0.970  \pm  0.007 \pm  0.035 $ &  $( 5.67 \pm  0.05  \stackrel{+ 0.94}{_{- 0.77} } )\times 10^{-2}$ \\
   158-184 &   $( 2.10 \pm  0.01 \stackrel{+ 0.36}{_{- 0.30} } )\times 10^{-2}$ &   $ 0.977  \pm  0.007 \pm  0.026 $ &  $( 2.05 \pm  0.02  \stackrel{+ 0.36}{_{- 0.30} } )\times 10^{-2}$ \\
   184-212 &   $( 7.47 \pm  0.05 \stackrel{+ 1.36}{_{- 1.16} } )\times 10^{-3}$ &   $ 0.983  \pm  0.007 \pm  0.019 $ &  $( 7.34 \pm  0.07  \stackrel{+ 1.35}{_{- 1.15} } )\times 10^{-3}$ \\
   212-244 &   $( 2.67 \pm  0.02 \stackrel{+ 0.52}{_{- 0.46} } )\times 10^{-3}$ &   $ 0.987  \pm  0.006 \pm  0.014 $ &  $( 2.63 \pm  0.02  \stackrel{+ 0.52}{_{- 0.45} } )\times 10^{-3}$ \\
   244-280 &   $( 8.88 \pm  0.10 \stackrel{+ 1.89}{_{- 1.69} } )\times 10^{-4}$ &   $ 0.990  \pm  0.006 \pm  0.009 $ &  $( 8.79 \pm  0.11  \stackrel{+ 1.87}{_{- 1.67} } )\times 10^{-4}$ \\
   280-318 &   $( 3.03 \pm  0.05 \stackrel{+ 0.72}{_{- 0.64} } )\times 10^{-4}$ &   $ 0.992  \pm  0.007 \pm  0.006 $ &  $( 3.01 \pm  0.06  \stackrel{+ 0.71}{_{- 0.63} } )\times 10^{-4}$ \\
   318-360 &   $( 9.53 \pm  0.27 \stackrel{+ 2.57}{_{- 2.21} } )\times 10^{-5}$ &   $ 0.993  \pm  0.006 \pm  0.004 $ &  $( 9.46 \pm  0.27  \stackrel{+ 2.55}{_{- 2.20} } )\times 10^{-5}$ \\
   360-404 &   $( 2.53 \pm  0.14 \stackrel{+ 0.79}{_{- 0.65} } )\times 10^{-5}$ &   $ 0.994  \pm  0.008 \pm  0.003 $ &  $( 2.51 \pm  0.14  \stackrel{+ 0.79}{_{- 0.64} } )\times 10^{-5}$ \\
   404-464 &   $( 6.34 \pm  0.61 \stackrel{+ 2.42}{_{- 1.81} } )\times 10^{-6}$ &   $ 0.994  \pm  0.010 \pm  0.002 $ &  $( 6.31 \pm  0.61  \stackrel{+ 2.40}{_{- 1.80} } )\times 10^{-6}$ \\
   464-530 &   $( 1.36 \pm  0.29 \stackrel{+ 0.65}{_{- 0.45} } )\times 10^{-6}$ &   $ 0.994  \pm  0.013 \pm  0.002 $ &  $( 1.36 \pm  0.29  \stackrel{+ 0.64}{_{- 0.44} } )\times 10^{-6}$ \\
   530-620 &   $( 2.78 \pm  1.24 \stackrel{+ 1.64}{_{- 1.11} } )\times 10^{-7}$ &   $ 0.994  \pm  0.008 \pm  0.003 $ &  $( 2.76 \pm  1.24  \stackrel{+ 1.63}{_{- 1.10} } )\times 10^{-7}$ \\

\end{tabular}
\end{ruledtabular}
\end{table*}
%%}

%%% >>>>>>>>>>>>>>>>>>
%%\arxivonly{
%%}
%%% >>>>>>>>>>>>>>>>>>

%%% >>>>>>>>>>>>>>>>>>
%%% >>>>>>>>>>>>>>>>>>

%%%%%%%%%%%%%%%%%
%%\bibliography{incjet_385_prdrc}

\begin{thebibliography}{00}
\bibitem{b:coordinate} We use a cylindrical coordinate system with the
  $z$ coordinate along the proton beam direction,
  %%KH and the origin at the center of the detector, 
  the azimuthal angle $\phi$, 
  and the polar angle $\theta$ usually expressed through the
  pseudorapidity $\eta=-\ln\tan(\theta/2)$.
  The rapidity $y$ is defined as $y=-1/2\ln((E+p_z)/(E-p_z))$.
\end{thebibliography}
%%%%%%%%%%%%%%%%%

\end{document}